\documentclass{aa}
\usepackage{epsfig}
\usepackage{graphics}
\usepackage{subfigure}
\usepackage{natbib}
\usepackage{amsmath}
\usepackage{enumerate}
\usepackage{threeparttable}
\usepackage{caption}

\begin{document}

\title{Gamma rays from red giant wind bubbles entering the jets of elliptical host blazars}

\author{N\'uria Torres-Alb\`a\inst{1} and Valent\'i Bosch-Ramon\inst{1}}
  
\institute{Departament de F\'{i}sica Qu\`antica i Astrof\'{i}sica, Institut de Ci\`encies del Cosmos (ICC), Universitat de Barcelona (IEEC-UB), Mart\'{i} i Franqu\`es 1, E08028 Barcelona, Spain}

\offprints{N. Torres-Albà \\ \email{ntorres@fqa.ub.es}}

\titlerunning{Gamma rays from wind bubbles}


\abstract
{Blazars in elliptical hosts have a population of red giants surrounding their jet. These stars can carry large wind-blown bubbles into the jets, leading to gamma-ray emission through bubble-jet interactions.}
{We study the interaction dynamics and the gamma-ray emission produced when the bubbles formed by red giant winds penetrate the jet of a blazar in an elliptical galaxy.}
{First, we characterized the masses and penetration rates of the red giant wind bubbles that enter the jet. Then, the dynamical evolution of these bubbles under the jet impact was analysed analytically and numerically, and the radiation losses of the particles accelerated in the interaction were characterised. Finally, the synchrotron and the inverse Compton contributions above $\sim 100$~MeV were estimated under different jet magnetic fields, powers, and Lorentz factors.}
{We find that an analytical dynamical model is a reasonable approximation for the red giant wind bubble-jet interaction. The radiation produced by these wind bubbles interacting with a jet can have a duty cycle of up to $\sim 1$. For realistic magnetic fields, gamma rays could be detectable from sources within the local universe, preferentially from those with high Lorentz factors ($\sim 10$), and this could be a relatively common phenomenon for these sources. For magnetic fields in equipartition with the jet power, and high acceleration rates, synchrotron gamma rays may be detectable even for modest Lorentz factors ($\sim 3$), but with a much lower duty cycle.}
{Blazars in elliptical galaxies within the local universe can produce detectable transient or persistent gamma-ray emission from red giant wind bubbles entering their jets.}
 
 
\maketitle

\section{Introduction}

Supermassive black holes, present in the innermost regions of galaxies, may accrete the material surrounding them, becoming active galactic nuclei (AGN). Some AGN produce collimated relativistic outflows, or jets \citep[e.g.][]{BegBla1984}, which propagate through the host galaxy. This propagation will inevitably lead to the jet interacting with a variety of obstacles including stars, gas, and dense clouds. These interactions may affect the jet dynamically \citep[e.g.][]{BlaKoe1979,WanWii2000,SutBic2007}. In particular, stars with high mass-loss rates may load the jet with enough matter to result in deceleration  \citep[e.g.][]{Kom1994,BowLea1996,HubBla2006,BosPer2012,PerMar2014,PerBos2017}.

The winds of stars interacting with AGN jets produce a double bow-shock structure in which particles can be accelerated to relativistic energies, possibly contributing to the jet's total non-thermal emission. Several works have explored this interaction and the resulting emission, both in the case of steady radiation and transient events \citep[e.g.][]{BedPro1997,BarAha2010,BarAha2012,KhaBar2013,AraBos2013,BedBan2015,WykHar2015,Bos2015,delBos2016,VieTor2017}, with results that often point towards possible detectability for nearby sources.

Previous works have considered persistent emission being generated from a whole population of stars, though in all cases they focus on the interaction that occurs within the jet once the star has already penetrated \citep{AraBos2013,WykHar2015,Bos2015,VieTor2017}. We refer to this stage of interaction as ``steady state''.

In this work, we focus on the possible mass-loading and emission generated at the moment when stars penetrate the jet, and the latter interacts with large ``bubbles'' of material formed by the collision between the stellar wind and the interstellar medium. \citet{PerBos2017} inferred possible significant non-thermal emission and mass-loading from this early stage in jet-star interaction, using 2D and 3D simulations of one single star with heavy mass loss. We use semi-analytical prescriptions to estimate if this is the case for a whole population of stars within a galaxy. 

We focus here on the study of blazar sources, as Doppler boosting is an important factor in enhancing the resulting emission. We consider only low-luminosity sources (i.e. $L_{\rm j}=10^{43}-10^{45}$~erg~s$^{-1}$), more abundant in the local universe. As recent studies show that the preferred hosts of blazars are late-type galaxies \citep[e.g.][]{ScaUrr2000,NilPur2003,FalPia2014,OlgLeo2016}, we model the red giant population within an elliptical bulge. A mostly phenomenological approach is adopted (with the exception of an illustrative numerical simulation), based on specific source knowledge and reference parameter values, as a first simplified step to explore the outcome of the wind bubble and jet interaction.

The paper is organized as follows. A description of the prescriptions used to characterize the stellar population is given in Sect.~\ref{population}. In Sect.~\ref{bubbles}, the properties of the stellar bubbles outside and immediately after penetrating the jet are described. The bubble evolution within the jet is described through analytical estimates and compared with simulation results in Sect. \ref{BubUp}. Then, the non-thermal emission generated by bubble-jet interactions is estimated in Sect.~\ref{appLnt}. Finally, the discussion is presented in Sect. \ref{Discussion}.

\section{Characterization of the stellar population in an elliptical galaxy}\label{population}

Elliptical galaxies contain large populations of red giants, which can have high mass-loss rates, in the range of $\dot{M} \sim 10^{-10} - 10^{-5}\,M_\odot$~yr$^{-1}$\citep{Rei1975}. We model the red giant population of any elliptical galaxy by taking as reference values those of M87, as its proximity allows for a precise study. 

\subsection{Stellar number density, mass-loss rate, and wind speed}\label{numberdensity}

Assuming a Kroupa initial mass function (IMF), and normalizing it to the total mass of stars ($M_{\rm T}$), we can estimate the number of stars within the bulge. We take as index for the IMF $x_1=-1.3$ for $0.1$ M$_\odot<M<0.5$ M$_\odot$ and $x_2=-2.3$ for $0.5$ M$_\odot<M<m_2$ \citep{Kro2001}, where $m_2$ is the mass of the stars exiting the red giant phase (i.e. the most massive stars present; see below). We estimate the total stellar mass adopting that contained within the galactic bulge in M87 \citep{GebTho2009}, knowing that the radius is $\sim 40''$  \citep{HarBir1999}, corresponding to a bulge radius of $R_{\rm b} \sim 3.1$ kpc, which we consider spherical.

From this point on, we follow the calculations in \citet{VieTor2017} to derive the mass of the red giants in the bulge of the galaxy assuming that all stars formed at the birth of the galaxy (i.e. no star formation extended in time), which yields $\sim 0.83 \, M_\odot$, and their number, which is $N_{\rm T}\sim 1.3 \times 10^9$. From \citet{GebTho2009}, we derive that the decay of the density with radial distance/jet height ($z$) can be approximated by $n_{\rm s}(z)\propto N_{\rm T}/z$ in the considered inner $\sim 40''$. 

We derive the mass-loss rate and wind speed of the red giant population exactly as in \citet{VieTor2017} for the particular case of M87. The result is a mass-loss rate that increases rapidly with the age of the red giant. Stars are thus modelled as a distribution that depends both on height and red-giant age (i.e. how deep into the red giant phase the star is), $n_{\rm s}(z, t_{\rm RG})$. For simplicity, we consider a stellar wind of  $v_{\rm w} = 10^7$ cm s$^{-1}$, and consider it constant during its evolution.

\subsection{Orbital velocities and penetration rate} \label{OrbitPR}

In order to study the collective emission and mass-loading generated by the whole population of stars as they penetrate the jet, we need to determine the frequency at which these events occur, that is, the penetration rate ($PR$). Knowing the orbital velocities of stars, one can estimate the penetration rate into the jet for the distance interval $(z,z+{\rm d}z)$ as ${\rm d}P(z, t_{\rm RG}) \simeq n_{\rm s} (z, t_{\rm RG}) v_{\rm orb}(z) R_{\rm j} (z) {\rm d}z$ \citep{KhaBar2013}, where $R_{\rm j}$ is the jet radius.

At low $z$ the stellar orbital movement is dominated by the central supermassive black hole, and thus stars orbit it following a Keplerian motion, with $v_{\rm orb}=\sqrt{G M_{\rm BH}/z}$. The gravitational influence of the black hole is dominant within a radius, or jet height, $z_{\rm g}={G M_{\rm BH}}/{\sigma ^2}$, where $\sigma$ is the stellar velocity dispersion of the bulge. At larger $z$, we consider the stars to move within the bulge at a constant velocity $\sigma$.

For the black-hole mass, we use the value derived by \citealp{GebAda2011}, $6.6 \pm 0.4 \times 10^9$ M$_{\odot}$. Dispersion velocity measurements decrease from $\sigma \sim 480$ km s$^{-1}$ near the nucleus (where the supermassive black hole is dominant) to $\sigma \sim 320$ km s$^{-1}$ at $\sim 40''$ \citep{GebAda2011}. We take the average value of $\sigma(z)$, $\sigma=360$ km s$^{-1}$, as the constant velocity for stars within the bulge, which gives $z_{\rm g} = 220$ pc ($\sim 2.5''$).

\section{Interaction with the jet}\label{bubbles}

\subsection{Stars outside the jet}\label{BubOut}

As a star moves outside the jet, the ram pressure generated by its stellar wind is in equilibrium with all external pressures, meaning
\begin{equation}\label{eq:windramp}
P_{\rm w} (\dot{M})= \rho_{\rm w} (\dot{M}) v_{\rm w}^2 \ = P_{\rm ext}
\end{equation}
The external pressures are given by the interstellar medium (ISM) thermal pressure $P_{\rm ISM}\approx 10^{-12}$~erg~cm$^{-2}$, and the orbital motion of the star through this medium, or ``orbital (ram) pressure'', is given by
\begin{equation}\label{eq:orbp}
P_{\rm orb}(z)=\rho_{\rm ISM} v_{\rm orb}(z)^2 \ ,
\end{equation}
where $\rho_{\rm ISM}$ is the ISM density. We fix the ISM density taking one hydrogen atom per cm$^{3}$, a typical value in the central regions of elliptical galaxies \citep{TanBeu2008}, throughout the entire bulge.

As the star approaches the jet, the jet lateral pressure $P_{\rm lat}$ may be larger than the pressures generated by the movement within the ISM:
\begin{equation}\label{eq:jetlatp}
P_{\rm lat} (z)=L_{\rm j} / c \pi z^2 \Gamma_{\rm j} ^2 \ ,
\end{equation}
where $L_{\rm j}$ and $\Gamma_{\rm j}$ are the jet luminosity and Lorentz factor, respectively \citep{BosBar2016}. Therefore, in close proximity with the jet, 
\begin{equation}\label{eq:Pressure}
P_{\rm ext} = {\rm max}(P_{\rm lat}, \ P_{\rm ISM} + P_{\rm orb}).
\end{equation}

We call $R_{\rm out}$ the distance from a star at which pressure equilibrium is reached outside the jet. At this distance, the colliding pressures generate a double bow shock in which both stellar-wind material and interstellar material are accumulated. This shocked layer surrounds the star in a bullet-like shape, with material gathering in the direction of the stellar movement toward the jet, and eventually escaping in the opposite direction.

We approximate this shocked region as a sphere of radius $R_{\rm out}$, in which an amount of material of mass $M_{\rm out}$ is contained, which can be estimated as
\begin{equation}\label{eq:BubbleOut}
R_{\rm out}^2 = \frac{\dot{M} v_{\rm w}}{4 \pi p_{\rm ext}}, \ \ \ M_{\rm out} = 4 \pi R_{\rm out}^3 \rho_{\rm w} (R_{\rm out}).
\end{equation}
We do not consider the mass accumulated within the shocked layer, which would be of the order of $\left( v_{\rm w}/ v_{\rm orb}\right)^2 M_{\rm out}\lesssim 0.1\,M_{\rm out}$. 

The dominant pressure component for most of the considered jet range is $P_{\rm orb}$, with $P_{\rm lat}$ being of comparable value at the base of the jet.

\subsection{Stars penetrating the jet}\label{BubIn}

As a star begins to penetrate the jet, the most external layers of the bubble it carries are hit by the jet ram pressure. This results in a shock that starts to propagate through those layers with a speed $c_s\approx \sqrt{{L_j}/{S_j c \rho_w(R)}} $. If the speed at which the shock propagates is sufficiently slow (i.e. the bubble is dense enough), part of the bubble material will penetrate the jet along with the star before the shock reaches the stagnation radius, $R_{\rm s}$. This will happen for all layers for which
\begin{equation}\label{eq:Rin}
c_s(R_{\rm in})<v_{\rm orb} \ , \ \ \ R_{\rm in}^2=\frac{v_{\rm orb}^2 \dot{M}}{4 \pi v_{\rm w} P_{\rm j}}\ ,
\end{equation}
where $P_{\rm j}$ is the jet pressure. This means layers with $R>R_{\rm in}$ will be expelled before the star fully penetrates the jet, and lost at the jet contact discontinuity (CD), while all layers within $R_{\rm in}$, with mass $M_{\rm in} = 4 \pi R_{\rm in}^3 \rho_{\rm w} (R_{\rm in})$, will manage to penetrate.

Once inside the jet, the shock will continue to propagate in the wind until it reaches the stagnation radius $R_{\rm s}$, the distance from the star where the wind and jet ram pressures are equal. There, a double bow shock is formed in which particles can be accelerated up to relativistic energies. In this work, we refer to this emission as steady state emission \citep[e.g.][]{VieTor2017}, as the jet-wind interaction process is continuous while the stars are inside the jet.\footnote{Instabilities produced at the jet-wind interaction region may actually lead to individual star-jet interaction variability \citep{delBos2016}.}

\begin{figure}[!ht]
\centering
\includegraphics[width=0.5\textwidth,keepaspectratio]{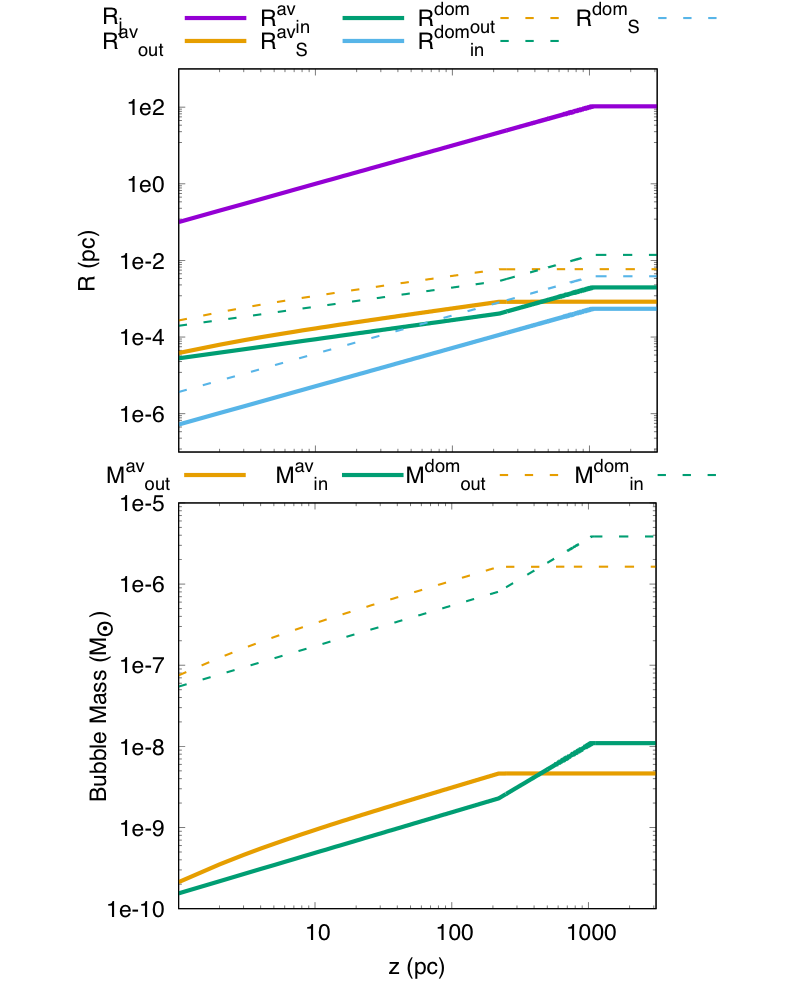}
\caption{Top: Radius of bubbles outside the jet (yellow), right after penetrating the jet (green) and at stagnation (blue) compared to the radius of the jet, as a function of jet height. Bottom: Mass within bubbles of radius $R_{\rm out}$ (yellow), and $R_{\rm in}$ (green), as a function of jet height. Parameters used are $L_{\rm j}=10^{44}$ erg s$^{-1}$, $\Gamma_{\rm j}=10$ and the orbital values given in Sect.~\ref{OrbitPR}. Values for both a red giant of average mass-loss rate (solid line) and a red giant that generates the dominant events (dashed line, see Sect.~\ref{BubRadiation}) are plotted.}
\label{fig:bubbles}
\end{figure}

Stellar material contained in the range $R_{\rm s}-R_{\rm in}$ is expelled within the jet as a blob \citep{BosPer2012,PerBos2017}. The typical initial sizes of the bubbles, compared with the jet radius, are shown in Fig.~\ref{fig:bubbles} for a red giant of average mass-loss rate ($\dot{M}=5.7\times 10^{-10}$ M$_{\odot}$ yr$^{-1}$), and for the red giant mass-loss rate that generates the dominant events ($\dot{M}_{\rm dom}=1.4\times 10^{-8}$ M$_{\odot}$ yr$^{-1}$, as calculated in Sect.~\ref{BubRadiation}).

For the particular set of parameters used to plot Fig.~\ref{fig:bubbles}, above $z \sim 400$~pc the size of the bubble inside the jet is shown as larger than the size of the bubble outside the jet. In such a case, the material introduced into the jet would be that contained within $R_{\rm s}-R_{\rm out}$. Another possibility, for a different set of parameters, is that $R_{\rm s}$ is larger than $R_{\rm in}$, but smaller than $R_{\rm out}$. In such a case, the star would lose any outer layers in the CD, and once inside the jet, the stellar wind termination region would expand up to the stagnation radius.

Figure \ref{fig:bubbles} shows the mass contained in the bubble outside and right after penetrating the jet, for the average and the event-dominant mass-loss rates. The jet is loaded with the external matter brought by the bubbles expelled by stars at penetration, though this mass-load rate is much lower than the jet  $\dot{M}_{\rm jet}=L_{\rm j}/\Gamma_{\rm j}c^2$, and therefore unlikely to result in a dynamical effect on the jet.

As has been studied for example by \citet{WykHar2015} and \citet{VieTor2017}, a population of high-mass stars in starburst galaxies can interact strongly with the jet. Young OB stars have stronger winds than red giants, with higher mass-loss rates and speeds. However, these stars have such fast winds that $v_{\rm w}>v_{\rm orb}$. Following Eqs. \ref{eq:BubbleOut}, when that condition is met, $R_{\rm s}>R_{\rm in}$. Therefore, the star penetrates the jet and the size of the interaction region surrounding the star actually increases, with no significant external material introduced into the jet.

\section{Bubble evolution within the jet}\label{BubUp}

\begin{figure*}[!t]
\centering
\includegraphics[width=0.95\textwidth,keepaspectratio]{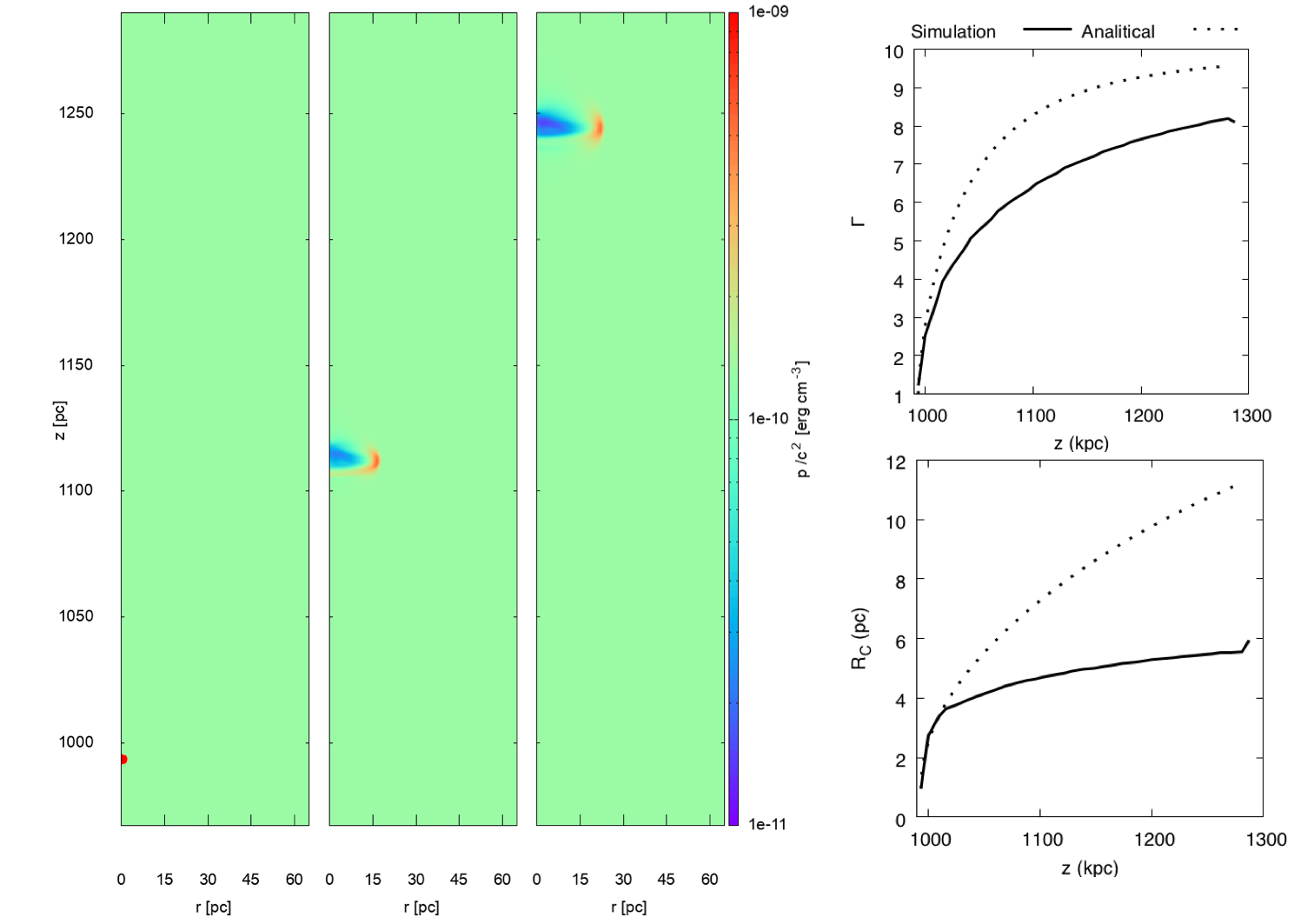} 
\caption{Left: Density maps obtained from a hydrodinamic simulation of the interaction between the jet and the penetrating bubble. Three snapshots are taken $\sim$0.4, 400 and 950 yr after penetration to illustrate bubble evolution. Top/Bottom right: Bubble Lorentz factor/Bubble radius as a function of jet height as calculated through the analytical estimate (dotted line) of \citet{BarAha2010,BarAha2012}, used to derive all results presented in this work, and compared to those obtained through our simulations (solid line).}
\label{fig:Simulation}
\end{figure*}

We consider a jet that initially expands with a conical geometry, launched close to the supermassive black hole in the centre of the galaxy. We consider that the jet recollimates, which we model as the jet becoming cylindrical:
\begin{equation}\label{eq:rJet}
R_{\rm{j}}(z) = \begin{cases} \theta\,z\ , & \mbox{if } z < z_{\rm eq} \\ Const, & \mbox{if } z \geq z_{\rm eq} \end{cases}\ ,
\end{equation}
when its pressure becomes equal to that of the ISM.

After the star penetrates the jet and a bubble of stellar material is expelled, this bubble evolves as a result of the interaction with the jet. We have adopted the analytical modelling of the evolution of a blob impacted by a jet developed by \cite{BarAha2010,BarAha2012} and \citet{KhaBar2013} \citep[for previous numerical simulations of this process, see e.g.][]{BosPer2012,PerBos2017}.

The shock produced by the impact of the jet ram pressure causes the material of the bubble to heat up, quickly expand, and accelerate, resulting in the bubble reaching relativistic speeds. This acceleration occurs on timescales of 
\begin{equation}\label{eq:t_acc}
t_{\rm acc} \simeq \begin{cases} z_{\rm 0}/\beta c , & \mbox{if } D < 1 \\ z_{\rm 0}/D \beta c , & \mbox{if } D > 1 \end{cases}\ , \ \ \  D \equiv \frac{P_{\rm j, 0} \pi R_{\rm b}^2 z_{\rm 0}}{4c^2 M_{\rm b} \Gamma_{\rm j}^3}\,,
\end{equation} 
where $D$ is a dimensionless parameter related to the jet luminosity and Lorentz factor and the mass of the bubble,  $R_{\rm b}$ is the radius of the bubble after acceleration \citep[see e.g.][]{BarAha2012}, and any subindex ``0'' refers to the value at $z_{\rm 0}$, the height at which the star penetrates. The value $D$ actually gives a comparison between bubble and jet mass on scales of $z_0$; that is, if $D>1$ ($<1$), the jet will (not) effectively accelerate the blob before it covers a distance $\sim z_0$. After the blob is accelerated, it is carried downstream (likely at least partially disrupted) until jet termination, at a height $H$ and at a speed $\sim \beta c$, thus on a timescale $t_{\rm esc} \sim (H-z_{\rm 0})/\beta c$.

This analytical approximation to the bubble evolution within the jet, which we use to calculate the results presented in Sect. \ref{appLnt}, can be compared to results obtained through numerical simulations. For that, we simulate here the bubble-jet interaction solving numerically the equations of relativistic hydrodynamics (RHD) assuming axisymmetry, a gas with constant adiabatic index $\bar{\gamma}=$4/3, and a dynamically negligible magnetic field. The RHD equations are solved using the Marquina flux formula \citep{DonMar1996,DonFon1998}; further details on the code can be found in \citet{delBos2016}. 

The resolution of the calculations is 1000 cells in the vertical direction, the z-axis, and 200 in the radial direction, the r-axis. Those numbers of cells correspond to a physical range of $z_{\rm grid}=1000 - 1300$ pc and $r_{\rm grid}=0 - 65$ pc. The number of cells was chosen such that the results did not change significantly when going to higher resolution. The boundary conditions were set to inflow at $z_{\rm grid}=0$ with parallel stream lines (the jet is recollimated), reflection at $r_{\rm grid}=0$, and outflow in the remaining boundaries.

Figure~\ref{fig:Simulation} shows three density maps, illustrative snapshots of a simulation of the jet-bubble interaction at $\sim$0.4, 400, and 950 yr after bubble penetration inside the jet. Initial properties are: $L_{\rm j}=10^{44}$ erg $s^{-1}$, $\Gamma_{\rm j}=10$, $M_{\rm b}=2 \times 10^{-6}$ M$_{\odot}$, and $r_{\rm b,0}=1.5$~pc. The simulation starts once the cloud has already expanded significantly, moving with a Lorentz factor of 2, in the relativistic regime \citep[e.g.][]{BarAha2012,KhaBar2013}. We focus only on the relativistic regime, as starting from its actual initial radius would require $10^2-100^2$ times more cells and thus huge computing costs. As seen in the figure, the shocked bubble evolves upstream of the jet; most of the bubble expansion occurs at early times, and despite partial disruption the structure keeps its integrity.  

In Fig.~\ref{fig:Simulation} we also plot the bubble radius and Lorentz factor as a function of jet height obtained from the simulation, and compare it to the results of the analytical calculation. The simulated evolution is slightly slower than analytically predicted. We note, however, that while the Lorentz factor of the bubble in the simulation reaches a value of $\sim 8$, it should eventually reach the jet Lorentz factor at higher $z$. The chosen duration of the simulation of $\sim 1000$~yr was adopted as a trade-off between a moderated computational time and illustrative effectiveness.

A noticeable difference in Fig. \ref{fig:Simulation} between simulated and analytical results is bubble radius, computed as the mass-averaged cylindrical radius to facilitate comparison. The radius evolves more slowly in the simulation, reaching a final value of a factor $\lesssim 2$ smaller than the analytical estimate. We expect nevertheless more convergence at larger $z$ values.

We show thus that the analytical first-order approximation to the bubble evolution used in this work, adopted to derive the results presented in Table \ref{table:results}, is reasonably accurate in the relativistic regime. While instabilities might be important, once the shocked bubble reaches relativistic speeds, its evolution is similar to the analytic predictions, even if disrupted as a filamentary fragmented structure mixed with shocked jet material. The reason is that the transversal expansion of the shocked bubble material is slowed down in the laboratory frame by flow-frame time dilation, and in the longitudinal direction by very similar bottom and the top speeds of the shocked structure in the laboratory frame. 

Therefore, from the simulation results, we find it reasonable at this stage to adopt a final constant bubble radius, and a final bubble speed close to that of the jet if $D>1$. For the bubbles that dominate the non-thermal emission (see Sect. \ref{BubRadiation}), this condition is fulfilled.

A more complete simulation, including earlier and later stages of the bubble evolution, is planned for future work.

\section{Radiated non-thermal power}\label{appLnt}

Particles are assumed to be efficiently accelerated at the interaction between the jet and the bubble. The non-thermal energy contained in these particles is uncertain, although we focus here on the case in which the energy budget is significant and detectable radiation may be produced. The specific acceleration mechanism is not considered, and it is just assumed that some fraction of the shocked jet luminosity goes to non-thermal particles.
 
 \begin{table*}[ht]
    \caption[]{Results.}
   	\label{table:results}
   	\centering
	\small
\begin{tabular}{lcccccc}
\hline\hline \\[-0.25cm]
$L_{\rm j}$ [erg s$^{-1}$]& $10^{43}$& $10^{44}$& $10^{45}$& $10^{43}$& $10^{44}$& $10^{45}$ \\
$\Gamma$ & 3 & 3 & 3 & 10 & 10 & 10 \\ [0.01cm]
\hline \hline \\[-0.25cm]
\multicolumn{3}{l}{Inverse Compton: $E'_{\rm e}=E'_{\rm IC}$ $B=0.1 B_{\rm eq}$}\\
\hline \hline \\[-0.25cm]
$L_{\rm st}$ [erg s$^{-1}$] & $6.5 \times 10^{39}$	& $1.7 \times 10^{40}$ & $1.7 \times 10^{40}$ & $2.5 \times 10^{39}$ & $6.9 \times 10^{39}$ & $1.8 \times 10^{40}$ \\ 
$L_{\rm pop}$ [erg s$^{-1}$]	& $1.1 \times 10^{38}$& $6.5 \times 10^{38}$ & $8.4 \times 10^{38}$ & $3.7 \times 10^{40}$ & $4.7 \times 10^{40}$ & $3.4 \times 10^{40}$\\ [0.01cm]
\hline \\[-0.25cm] 
$L_{\rm dom}$ [erg s$^{-1}$]  & $1.4 \times 10^{38}$& $8.8 \times 10^{37}$ & $5.5\times 10^{37}$ & $3.5 \times 10^{40}$ & $3.1 \times 10^{40}$ & $2.5 \times 10^{40}$\\
$<M_{\rm b}>$ [$M_\odot$]& $1.6 \times 10^{-6}$ & $1.7 \times 10^{-6}$ & $1.5 \times 10^{-6}$ & $1.7 \times 10^{-6}$& $1.7 \times 10^{-6}$	& $1.5 \times 10^{-6}$ 	 \\
$<z_{\rm 0}>$ [kpc]  & 0.33& 1.0 & 1.6 & 0.67 & 1.1 & 1.6 \\
$PR$ [yr$^{-1}$]  & $8.3 \times 10^{-3}$	& $2.6 \times 10^{-2}$ & $4.0 \times 10^{-2}$ & $8.6 \times 10^{-3}$& $2.7 \times 10^{-2}$ & $4.0 \times 10^{-2}$ \\
$t_{\rm obs}$ [yr]  & $54$ & $170$ & $270$ & $40$ & $33$ & $24$ \\
\hline \hline \\[-0.25cm]
\multicolumn{3}{l}{Synchrotron: $E'_{\rm e}=E'_{\rm Sy}$, $B=B_{\rm eq}$}\\
\hline \hline \\[-0.25cm]
$L_{\rm st}$ [erg s$^{-1}$] & $7.5 \times 10^{41}$	& $1.9 \times 10^{42}$ & $1.9 \times 10^{42}$ & $1.3 \times 10^{41}$ & $7.4 \times 10^{41}$ & $2.5 \times 10^{42}$ \\
$L_{\rm pop}$ [erg s$^{-1}$]	& $1.4 \times 10^{40}$& $4.0 \times 10^{40}$ & $6.7 \times 10^{40}$ & $1.1 \times 10^{42}$ & $2.0 \times 10^{42}$ & $2.5 \times 10^{42}$\\  [0.01cm]
\hline \\[-0.25cm]  
$L_{\rm peak}$ [erg s$^{-1}$]  & $5.8 \times 10^{41}$& $5.8 \times 10^{42}$ & $5.8 \times 10^{43}$ & $6.3 \times 10^{42}$ & $6.3 \times 10^{43}$ & $6.4 \times 10^{44}$\\
$<M_{\rm b}>$ [$M_\odot$]& $1.5 \times 10^{-6}$ & $1.7 \times 10^{-6}$ & $1.6 \times 10^{-6}$ & $1.7 \times 10^{-6}$& $1.7 \times 10^{-6}$	& $1.6 \times 10^{-6}$ 	 \\
$<z_{\rm 0}>$ [kpc]  & 0.26& 0.88 & 1.9 & 0.91 & 1.3 & 1.9 \\
$PR$ [yr$^{-1}$]  & $5.7 \times 10^{-3}$	& $2.0 \times 10^{-2}$ & $3.8 \times 10^{-2}$ & $7.4 \times 10^{-3}$& $2.3 \times 10^{-2}$ & $3.4 \times 10^{-2}$ \\
$t_{\rm peak}$ [yr]  & 1.3& 0.14 & 0.014 & 4.8 & 0.48 & 0.046 \\
\hline \hline \\[-0.25cm]
\multicolumn{3}{l}{Synchrotron: $E'_{\rm e}=E'_{\rm Sy}$, $B=B_{\rm min}$}\\
\hline \hline \\[-0.25cm]
$B_{\rm min}$ [$B_{\rm eq}$]& $0.02$ & $0.01$ & $0.01$ & $0.04$& $0.03$	& $0.02$ 	 \\
$L_{\rm pop}$ [erg s$^{-1}$]	& $2.1 \times 10^{40}$& $6.0 \times 10^{40}$ & $1.4 \times 10^{41}$ & $1.7 \times 10^{42}$ & $1.8 \times 10^{42}$ & $1.4 \times 10^{42}$\\ [0.01cm]
\hline \\[-0.25cm] 
$L_{\rm dom}$ [erg s$^{-1}$]  & $6.9 \times 10^{40}$& $2.1 \times 10^{40}$ & $4.3 \times 10^{40}$ & $1.8 \times 10^{42}$ & $1.1 \times 10^{42}$ & $1.9\times 10^{42}$\\
$t_{\rm obs}$ [yr]  & $13$ & $45$ & $32$ & $38$ & $65$ & $10$ \\

\hline  \\[0.005cm]

\end{tabular}
\begin{tablenotes}
{\footnotesize
\item {\bf Notes:} Results for the emission of the whole population of red giants interacting with the jet in steady state ($L_{\rm st}$) and through injecting bubbles at penetration ($L_{\rm pop}$), respectively; $L_{\rm dom}$ is the luminosity of the typical bubble event that dominates the aforementioned emission, and below it the characteristics of this event are given. For six different jet configurations, we list all results in three cases: inverse Compton emission of electrons with energy $E'_{\rm IC}$ for $B=0.1B_{\rm eq}$, and synchrotron emission at 100~MeV for $B_{\rm eq}$ and $B_{\rm min}$. The non-listed characteristics of the dominant event in the case of synchrotron emission with $B_{\rm min}$ are the same as those of the emission with $B_{\rm eq}$, for each jet configuration. \newline All the luminosities and energies scale with $\eta/0.1\le 10$.
\newline The luminosities and times are as seen by the observer.\\}
\end{tablenotes}
\end{table*}

\subsection{Energy losses}

We assume that electrons (and positrons) are the dominant non-thermal emitting particles. Accelerated protons may also be present in the jet, although on the scales of interest significant radiation losses are not expected for these particles \citep[see however, e.g.][]{Aha2000}. For electrons, we considered the radiative losses via inverse Compton (IC) and synchrotron emission. Regarding non-radiative losses, we considered adiabatic losses for the conical jet, and none after recollimation (but escape when the bubble reaches $H$).

We computed the IC losses as in \citet{BosKha2009} \citep[see also][]{KhaAha2014}. Their approximation is valid for a Planck distribution of target photons of temperature $T$, and must be renormalized to the energy density of the considered target photon field. We consider the photon field in an elliptical galaxy as derived by \citet{VieTor2017}.

We also considered synchrotron losses \citep[e.g.][]{Lon1981} when taking $B=B_{\rm eq}$, a magnetic field of equivalent energy density to that of the jet, meaning 
\begin{equation}
\frac{B^{2}_0}{4 \pi} =  \frac{L_{\rm{j}}}{\pi R_{\rm{j}}(z_{0})^2 c}\,,
\end{equation}
with the field depending on height as
\begin{equation}
B(z) = B_0 \left(\frac{z_0}{z}\right)^{2}\,,
\end{equation}
where $B_0 = B(z_0)$. We discuss the effects on the synchrotron radiation of considering a lower magnetic field in Sect.~\ref{LowB}.

\subsection{Radiation}\label{BubRadiation}

For all combinations of $L_{\rm j}= 10^{43},\ 10^{44},\ 10^{45}$, and $\Gamma_{\rm j}=3,\ 10$, we computed the average luminosity of the whole population of bubbles penetrating the jet (as seen by the observer), $L_{\rm pop}$. Results are presented in Table~\ref{table:results}. All parameters except for jet luminosity, $L_{\rm j}$, and Lorentz factor, $\Gamma_{\rm j}$, are fixed as described in Sects.~\ref{population} and \ref{bubbles}. 

We evaluated the emitted non-thermal radiation at two characteristic electron energies: first, for IC emission, at the Thompson-Klein Nishina transition energy, at $E'_{\rm IC}= (m_{\rm e}c^2)^2/kT\Gamma$, where the maximum of IC emission is expected for reasonable electron energy distributions, falling in the gigaelectronvolt-Teraelectronvolt range; secondly, for synchrotron emission, for the electrons that generate synchrotron 100~MeV photons as seen by the observer, at $E'_{\rm sy}$.

We assumed that at most a fraction $\eta$ of the energy acquired by jet acceleration can be radiated\footnote{The bubble acquires in the laboratory frame an energy $\sim \Gamma_{\rm j}M_{\rm b}c^2$ due to jet acceleration.}, fixed to 0.1 throughout this work.  From first principles, it is not possible to derive the value of $\eta$, but the adopted choice maximizes the predicted gamma-ray emission without assuming a full non-thermal conversion of the available energy. All the predicted luminosities thus scale with $\eta/0.1\le 10$. 

We estimated the typical properties of the bubbles that contribute the most to the overall luminosity as
\begin{equation}\label{Averages}
<A> = \frac{\int A(z, t_{\rm RG})L_{\rm pop}(z,t_{\rm RG})\ {\rm d}z\ {\rm d}t_{\rm RG}}{L_{\rm pop}(z,t_{\rm RG})\ {\rm d}z\ {\rm d}t_{\rm RG}}\,,
\end{equation}
where $A$ denotes the quantity we are interested in evaluating (i.e. $M_{\rm b}$ and $z_{\rm 0}$). We refer to events with these properties as the ``dominant'' events, and list their characteristics in Table~\ref{table:results}. Their penetration rate (i.e. rate at which events occur) along with peak luminosity of the event, $L_{\rm dom}$ or $L_{\rm peak}$, and typical duration of the peak, $t_{\rm obs}$ or $t_{\rm peak}$, are also listed. We note that $L_{\rm dom}$ is the average luminosity of a dominant event along its evolution within the jet; in reality, the luminosity is a function of jet height (see Eq. \ref{eq:Lum1BIC}), larger at $z_0$, where the star penetrates, and becomes progressively dimmer as the radiative cooling efficiency diminishes.

We also compared the emission generated through this interaction mechanism to that generated in steady state, $L_{\rm st}$, by the same population of red giants (see Sect.~\ref{bubbles}). The apparent non-thermal luminosity per unit volume at a height $z$ due to jet-star interactions in steady state is
\begin{equation}\label{eq:appLnt}
\frac{{\rm d}L_{\rm st}(z)}{{\rm d}V} = \eta L_{\rm j}  F_{\rm rad}(z)  \frac{\delta_{\rm j}^3}{\Gamma_{\rm j}^3} \int{  \left\langle \frac{S_{\rm s}(z, t)}{S_{\rm j}(z)}\right\rangle n_{\rm s}(z,t) {\rm d}t}\,,
\end{equation}
where $<{S_{\rm s} (m,t)}/{S_{\rm j}}>$ is the time-averaged fraction of jet area intercepted by one stellar interaction. 

\subsubsection{Inverse Compton emission}\label{ICemission}

At electron energies $E'_{\rm IC}$, for all jet parameters considered, $t'_{\rm IC} \sim 10^{12}-10^{13}$~s. This is much larger than the characteristic acceleration time of any considered bubble, typically in the range $t'_{\rm acc}\sim 10^6 - 10^9$~s. Therefore, the IC emission at the energy of interest will last until long after the bubble has been accelerated.

As the bubble propagates downstream in the jet, the former radiates an IC luminosity $L'_{\rm b, IC}\sim E'_{\rm b,max}/t'_{\rm IC}(z)$ in the flow frame, where $E'_{\rm b}$ is the total energy emitted by one single bubble in the same frame during its evolution
\begin{equation}\label{eq:Lum1BIC}
E'_{\rm b}(M_{\rm b}, z_{\rm 0})= \int_{z_{\rm 0}}^{z_{\rm max}} \eta \frac{M_{\rm b} c^2}{t'_{\rm IC}(z)}\frac{{\rm d}z}{\Gamma \beta c}\,;
\end{equation}
$z_{\rm max}$ is the maximum height the bubble reaches before losing all energy, or the jet maximum height, $H$. This approximation implicitly assumes that the energy distribution of the non-thermal electrons is $\propto E^{-2}$ (see Sect.~\ref{corr}). 

\begin{figure}[!ht]
\centering
\includegraphics[width=0.45\textwidth,keepaspectratio]{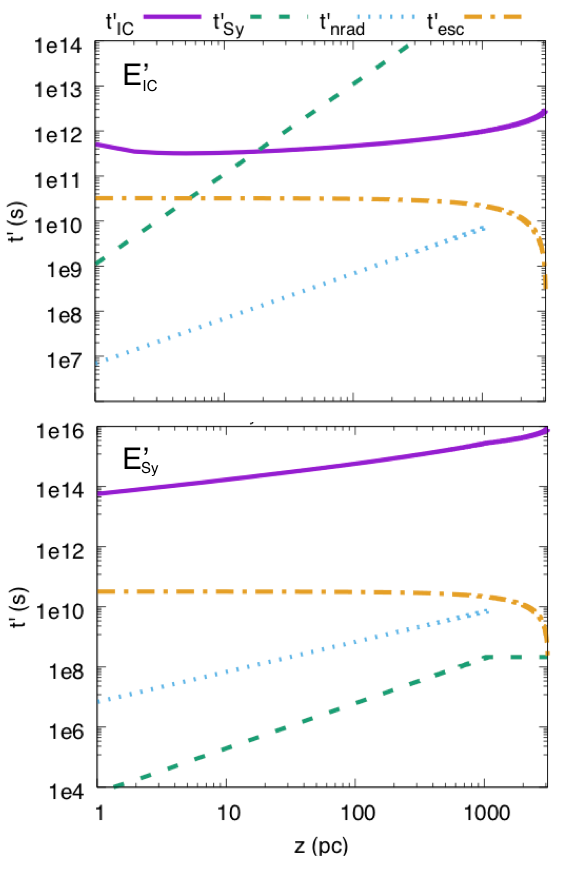}
\caption{Inverse Compton and synchrotron loss times for a jet with $L_j=10^{44}$ erg s$^{-1}$, $\Gamma_{\rm j}=10$, and (top) $B=0.1 B_{\rm eq}$, evaluated at $E'_{\rm IC}$; (bottom) $B=B_{\rm eq}$, evaluated at $E'_{\rm Sy}$.  The time for the bubble to reach the jet termination height, and the time of adiabatic losses, are also plotted.}
\label{fig:losses}
\end{figure}

Characteristic times for energy losses evaluated at $E'_{\rm IC}$, as well as $t_{\rm esc}$, are plotted in Fig.~\ref{fig:losses}. The energy losses are dominated by adiabatic expansion up to the jet recollimation point. At higher $z$, the shortest timescale is $t'_{\rm esc}$, meaning the bubbles exit the jet before emitting (mostly through IC) all of their available energy. 

We note that, contrary to the study of synchrotron gamma-ray emission below, IC energy losses are computed taking $B=0.1B_{\rm eq}$, which is a reasonable value for a jet with power dominated by its kinetic energy. However, since the energy losses at $E'_{\rm IC}$ are dominated by non-radiative processes, considering more (or less) intense magnetic fields does not significantly affect the results.

We can estimate the luminosity detected by an observer, generated by the whole bubble population, averaged in time as
\begin{equation}\label{eq:LumIC}
L_{\rm pop}= \delta_{\rm j}^3 \int_{z_{\rm BH}}^{H} E'_{\rm b}(m,z) PR(m,z) \ {\rm d}m \ {\rm d}z \,,
\end{equation}
where $\delta_{\rm{j}}$ is the Doppler boosting factor. This is a valid description as long as there is more than one dominant bubble simultaneously emitting\footnote{The Doppler-boosted luminosity of one single bubble is $L^{app}_{\rm b}= \delta^4 L'_{\rm b}$. If more than one star is within the jet emitting at different heights, the photons emitted by the distribution of sources at the same time in $K'$ are not observed simultaneously in $K$, leading to $\delta^4\rightarrow \delta^3/\Gamma$ \citep[see][]{SikMad1997}. Thus, if insufficient events take place at the same time,
the apparent luminosity of the whole population can be reduced to that of one single bubble.}, meaning that the source duty cycle is larger than 1. In addition, the emission generated by the whole population, $L_{\rm pop}$, as seen by the observer will be larger than the emission produced by one single, dominant event, $L_{\rm dom}$.

The predicted apparent luminosities of the IC emission are a few times $10^{40}$~erg~s$^{-1}$. These values are of the order of the steady state emission generated by the whole stellar population within the jet.

\subsubsection{Synchrotron emission}\label{Syemission}

If we evaluate the energy losses at $E'_{\rm Sy}$, $t'_{\rm Sy}$ is dominant at all heights for $B'=B'_{\rm eq}$ (see Fig.~\ref{fig:losses}). The synchrotron emission is so intense that the characteristic emission time is lower than the acceleration time of the bubble, $t'_{\rm acc}$. Therefore, electrons of energy $E'_{\rm Sy}$ immediately radiate via synchrotron emission the energy acquired via particle acceleration triggered by the jet-bubble interaction. Electrons in the flow frame are accelerated at a rate $\xi qB'c$, where $\xi$, a free parameter representing the acceleration efficiency, is fixed to 0.1. Such an efficiency is sufficient to produce observable 100~MeV synchrotron photons, although for $\xi$ well below 0.1 synchrotron emission would not reach gamma-ray energies (X-rays of luminosities similar to those of IC in Table~\ref{table:results} could still be produced). 

Synchrotron 100~MeV photons are produced when a single bubble enters the jet and is accelerated, following a flare-like lightcurve \citep[see][]{BarAha2012}. From \cite{KhaBar2013}, the apparent total energy emitted by a bubble in the synchrotron fast cooling regime is:
\begin{equation}\label{eq:Energy1BSy}
E_{\rm b} (M_{\rm b},z) = \eta F_{\rm rad} \bar{F_{\rm e}}M_{\rm b} c^2 \delta_j^3 \,,
\end{equation}
where we take $\bar{F_{\rm e}}=0.2$ \citep{BarAha2010}, and $F_{\rm rad}$ is the efficiency with which the particle loses energy through synchrotron radiation. Doppler boosting is already accounted for in Eq.~\ref{eq:Energy1BSy} for the energy radiated by one single bubble.

Unlike in the case of the IC emission, synchrotron $L_{\rm peak}$ is 1--3 orders of magnitude higher than the average luminosity $L_{\rm pop}$, with
\begin{equation}\label{eq:Lum1BSy}
L_{\rm peak} (M_{\rm b},z) = \eta F_{\rm rad} c F_{\rm e,max} P_{\rm 0} \delta_{\rm j}^2 \pi r^2_{\rm b}\,,
\end{equation}
where we take $F_{\rm e,max}=0.4$ \citep{BarAha2010}, and $r_{\rm b}$ is the radius of the bubble once it has expanded and reached relativistic velocities \citep{KhaBar2013}. The apparent duration of this intense emission can be roughly estimated as $t_{\rm peak} = E_{\rm b}/L_{\rm b}$, which shows that the emission is highly variable, like intense, short flares occurring a few times per century or millennia. 

For the high $\xi$ adopted, the synchrotron 100 MeV emission could reach luminosities of $\sim 10^{44}$~erg~s$^{-1}$ for the most powerful jets considered, even under low jet Lorentz factors. As the emission is intense but short, we would have duty cycles of $PR \cdot t_{\rm obs} \sim 10^{-4}-10^{-2}$, depending on $L_{\rm j}$ and $\Gamma_{\rm j}$.
 
\subsubsection{Synchrotron emission at low magnetic fields}\label{LowB}

Synchrotron emission at $\sim 100$~MeV is highly dependent on the value of the magnetic field: high magnetic fields yield an intense emission radiated in a very short amount of time, yielding a low duty cycle. 

In order to maximize the duty cycle, we take the lowest possible value of the magnetic field, $B=B_{\rm min}$, for which electrons with $E'_{\rm Sy}$ still cool more efficiently through synchrotron emission than through non-radiative losses. Looking at Fig.~\ref{fig:losses}, this would result in values of $t'_{\rm Sy}$ just below $t'_{\rm ad}$ and/or $t'_{\rm esc}$. The obtained magnetic field values, $B_{\rm min}$, are listed in Table~\ref{table:results}. For these values of $B$, luminosity is up to $\sim 10^{42}$~erg~s$^{-1}$, but the duty cycle can be increased up to $PR \cdot t_{\rm obs} >1$.

We note that in this case with $B=B_{\rm min}$, $t'_{\rm sy} > t'_{\rm acc}$, meaning that the synchrotron emission is computed as described in Sect.~\ref{ICemission}, that is, as is done for inverse Compton radiation. 

\section{Discussion}\label{Discussion}

The dynamical evolution of a bubble of stellar material expelled within the jet is semi-quantitatively well-described by the analytical estimates used in Sect.~\ref{BubUp}, and only differs from simulations by a moderate numerical factor. Our numerical estimates on gamma-ray energy production are not significantly affected by these differences as long as the bubble Lorentz factor eventually reaches $\sim \Gamma_{\rm j}$. Only in the case of intense Synchrotron emission at high magnetic fields, could there be a small reduction of luminosity, of a factor of a few, if indeed the bubble radius were overestimated in the long run.

Results presented in Table~\ref{table:results} indicate that the emission generated by stars penetrating the jet can be relatively persistent at high energies, through either inverse Compton or through synchrotron emission in the case of low magnetic fields. With large magnetic fields, emission at 100~MeV could be dominated by bright and infrequent events on top of the persistent, lower IC radiation.

The steady state emission of the whole population is unlikely to be detectable. We note that the similar values of $L_{\rm st}$ for all the explored jet configurations are due to the fact that it does not strongly depend on either $\Gamma_{\rm j}$ or $L_{\rm j}$. The small differences are due to jet geometry, which influences the amount of stars within the jet (e.g. a jet of $L_{\rm j}=10^{43}$~erg~s$^{-1}$ recollimates at low heights), or due to differences in $F_{\rm rad}$ (e.g. under high $B$ values, for $L_{\rm j}=10^{45}$ the impact of synchrotron losses on electrons at $E'_{\rm IC}$ is noticeable). 

Inverse Compton emission of the bubbles at $E'_{\rm IC}$ seems difficult to detect for the explored jet configurations unless $\eta\rightarrow 1$, or for a very nearby source. This is because non-radiative losses, or even synchrotron losses, dominate the process.

Synchrotron emission at $E'_{\rm Sy}$ with high magnetic fields can result in bright, detectable flares, although the high luminosity implies a short duration. If we consider lower magnetic fields, persistent emission can be achieved, and luminosities of the order of $10^{42}$~erg~s$^{-1}$ for $\Gamma_{\rm j}=10$.

There are some factors that may easily increase the radiation in the scenario studied here. For instance, for significantly lower values of $\rho_{\rm ISM}$ at $z\sim$~kpc, say $\gtrsim 0.1~$cm$^{-3}$, the emission energetics would grow by a factor of several due to the associated larger bubble mass (see Sect.~\ref{BubIn}), limited now by $M_{\rm in}$ (see Fig.~\ref{fig:bubbles}). Furthermore, a younger, more massive red giant population, or the sporadic presence of an asymptotic giant branch (AGB) star within the jet, could also enhance the expected emission.

\subsection{Younger red giant populations and AGB stars}

Our results show that the emission produced by wind bubbles as they penetrate the jet is generally dominated by the most evolved stars within our modelled population (with $\dot{M} \sim 1.4 \times 10^{-8}$~M$_{\odot}$~yr$^{-1}$). The mass of the bubbles scales with mass-loss rate as $\propto \dot{M}^{3/2}$, and the luminosity emitted by the bubble interacting with the jet is proportional to its mass (except for the case of synchrotron emission of 100~MeV photons at $B_{\rm eq}$). Therefore, our results are scalable with mass-loss rate. 

If we had a population of red giants of $M_{\rm RG} = 1.5$ M$_{\odot}$ with a total mass comparable to that of the Galaxy, it would imply an increase of luminosity of a factor of $\sim 20$ in the case of low-B synchrotron emission, and of a factor of $\sim 60$ in the case of Inverse Compton emission. In the case of high-B emission, emission depends on the final radius of the bubble, and it would increase by a factor of $\sim 6$. In all cases, considering these less abundant stars would lead to a decrease of event occurences of $\sim 0.4$.

We note that in the rare occasion an AGB star penetrates an AGN jet, its wind bubble can inject into the latter up to $\sim 10^{-2}$ M$_\odot$. This could potentially lead to a long duration event with IC luminosity $\sim 10^{44}$~erg~s$^{-1}$.

\subsection{Caveats of the radiation estimates}\label{corr}

In this work we estimate the energy radiated by accelerated particles at a given energy, $E'_{\rm IC,sy}$, where the maximum of emission is expected to be. In assuming that all the available energy that goes into particle acceleration is radiated or lost at the mentioned energy, we are overestimating the overall emission. Depending on the energy distribution of the particles, this simple approximation can differ from a more precise calculation. In the case of inverse Compton emission, we expect this overestimation to be of a factor of a few, and in the case of synchrotron emission, of up to one order of magnitude (\citealp{VieTor2017}; Vieyro et al. in prep). This has to be taken into account when reading the luminosities in Table~\ref{table:results}. The approximations adopted are valid for electron energy distributions $\propto E^2$, typical for astrophysical sources. Harder electron distributions would still lead to similar results to those obtained but for very extreme cases, while steeper distributions would imply an even higher overestimate of the gamma-ray luminosity. 

The radiation estimates were focused on gamma rays and a broad band study would require more detailed modelling. Nevertheless, it is worth exploring in the future the radio and X-ray synchrotron emission in the jet-bubble interactions. 

\section{Summary}\label{conclusion} 

In this work we have studied the gamma-ray emission produced when the bubbles formed by red giant winds penetrate the jet of a blazar in an elliptical galaxy, and described their dynamical evolution both analytically and through one illustrative simulation. We have shown that the analytical approximation is reasonably accurate at the present stage in the relativistic regime. We have found that the gamma rays produced by the red giant wind bubbles interacting with jets may reach detectable levels if Lorentz factors are high, non-thermal particles are generated very efficiently, and particle acceleration takes place at very high rates (to produce synchrotron 100~MeV photons). This predicted emission could be higher under the presence of an important population of younger red giants, in the rare event an AGB star enters the jet, or for relatively low values of $\rho_{\rm ISM}$. Unless $B=B_{\rm eq}$, duty cycles are not far below one, and in some cases, a few dominant bubbles could contribute simultaneously to the gamma-ray emission. For $B=B_{\rm eq}$, short bright synchrotron 100~MeV flares may be detected, with year or sub-year scale duration. 
Most of the known blazars in the local universe are hosted by elliptical galaxies. Therefore, provided that electrons are efficiently accelerated in bubble-jet interactions, it is plausible that some persistent or transient gamma-ray emission detected from the nearest blazars could originate in events like those described here. 

\section*{Acknowledgments}
We want to thank the anonymous referee for constructive and useful comments.
We acknowledge support by the Spanish Ministerio de Econom\'{i}a y Competitividad (MINECO/FEDER, UE) under grants AYA2013-47447-C3-1-P, AYA2016-76012-C3-1-P, with partial support by the European Regional Development Fund (ERDF/FEDER), MDM-2014-0369 of ICCUB (Unidad de Excelencia `Mar\'{i}a de Maeztu'), and the Catalan DEC grant 2017 SGR 643. N.T-A. acknowledges support from MINECO through FPU14/04887 grant. 
\bibliographystyle{aa}
\bibliography{../Bibliografia/Referencies}   


\end{document}